\documentstyle[12pt,aasms4]{article}
\newcommand{\Msol}{M$_\odot$}
\begin{document}

%
%

\title{Simplified treatment of the radiative transfer problem in expanding
envelopes}
\author{
J. G\'{o}mez-Gomar \altaffilmark{1}
\and J. Isern \altaffilmark{1}}
\altaffiltext{1}{Centre d'Estudis Avan\c{c}ats de Blanes (CSIC), Blanes, 
Cam\'{\i} de Santa Barbara s/n, C.P. 17300, Girona, Spain.}

\received{}
\accepted{}

\slugcomment{\underline{Submitted to}: \apj \underline{Version}: 
\today}

\begin{abstract}
In this paper we study the application of a simplified method to solve         
the dynamic radiative transfer problem in expanding envelopes. 
 The method, which requires a computational effort similar to that of the
 diffusion approximation,  is based on the use of a generalization of the 
Eddington closure relationship  allowing the inclusion of scattering and relativistic
corrections to $O \left ( v/c \right )$.
We apply this method to the calculation of light curves of type Ia supernovae,
 showing that it gives much more accurate results than the diffusion 
approximation, and that the latter is seriously in error when applied to
determine emergent flux and its spectral distribution.
 \end{abstract}

\section{Introduction}
 Radiation transfer plays a very important role in many
 astrophysical systems. In the case of expanding extended envelopes, like
those found on supernova or nova outbursts, this problem turns out to be
crucial since  the optically thin region  represents
a non-negligible part of the system. The evolution of these envelopes 
is  strongly influenced, or even dominated, by radiation and their 
detailed modeling is complex and  time consuming.

Very often the ``diffusion approximation'' is used to study these problems.
 (\cite{Ar79}, \cite{Ar80}, \cite{Ka94}).
  However, the basic hypotheses lying beneath 
this method, which is accurate in optically thick media, are no longer 
valid  in the envelopes considered here. First of all, the mean free path
of the photons and  the characteristic length of the envelope are of the
 same order. 
 As a consequence, the radiation field becomes anisotropic due to
the geometrical effects.  Secondly, the energy distribution of the radiation
 is not coupled to the local properties of  matter, specially if 
scattering is dominant, and the contribution of non-local radiation
to the continuum  makes inadequate the  assumption of thermal equilibrium
 between matter and radiation. 
 
To circumvent this problem the general moment equations (from which
 the diffusion approximation is derived) is used when accurate treatments
are required.
However, to solve the moment equations it is necessary to add
 a closure equation  to obtain a unique solution. This
 "closure relationship" is not known "a priori",
except in some particular cases  like those   where the
well known Eddington relationship
 $\left ( P_{\nu} / E_{\nu}  =1/3\right )$ holds. 
 Very often
the ``variable Eddington factor'' is used. In this case, the closure 
relationship is
 obtained by solving, from time to time, the complete transfer equation 
 in its static version. 
 The radiation moments are then computed from the intensities determined 
 in this way
  and the variable Eddington
 factor $\left (f_{\nu} =   P_{\nu} / E_{\nu} \right )$ is obtained 
(\cite{St92}, \cite{Ho93}). Although 
this method is quite efficient, it still demands an important computational effort
 in highly dynamical systems, where the transfer equation must be solved 
frequently.

  A  generalization of the Eddington 
relationship valid for static extended envelopes was developed by 
Simonneau (1979). The efficiency of
 this approximation was tested
 in static envelopes similar to those found in supernova
 outbursts
 (except for the velocity field ) in the cases where scattering
 was absent or dominant (Lopez, Simonneau  \& Isern 1987, Simonneau, Isern \& 
Lopez 1989). The main advantage of this relationship is that it can be 
easily obtained a priori. 
In this paper we test the  use of this relationship 
in the dynamic case and we show that, although the computational requirements
of this  method are similar to those of the 
the diffusion approximation,  the results are noticeably improved.

\section{The model}
As the test scenario we have chosen the semi-analytic model for a SNIa proposed
by
Schurmann (\cite{Sh83}). The expanding structure is composed by a 0.8 
\Msol{~} constant density core, surrounded by a 0.6 \Msol{~} 
envelope with a
density profile $\rho \propto r^{-7}$. The object is homologously
expanding with a total kinetic energy of 1.2 $\times 10^{51}$ ergs. The core 
contains 0.7 \Msol{~}
of totally burned matter ($\approx$ 100\% $^{56}$Ni just after the explosion),
 and a 0.1 \Msol{~} partially burned mantle, while the envelope is formed
by unburned C/O. The only source of energy  is the
radioactive decay of $^{56}$ Ni and $^{56}$ Co in the core. The energy
deposited by positrons and $\gamma$-rays produced in the decays is given
by an analytic fit to  Monte Carlo simulations performed by Colgate
and Petschek (1980) for very similar models. 

Three different total opacities have been used: $\chi=0.05\:\rho
$ $ cm^{-1}$,    $\chi=0.2\:\rho$ $cm^{-1}$  and 
pure free electron opacity (temperature and density dependent). 
In all the cases, the opacities are assumed to be independent of the
frequency.
The temperature dependent opacity presents a rather
realistic evolution but, since no line contributions were considered, it
 might fall to too low values at late phases. To avoid that, a lowest 
value of $\chi=0.005 \rho$ cm$^{-1}$ was imposed. Different parameterizations
were also adopted for the fraction of  pure absorption: $\epsilon$=1, 0.1 and
 0.01.

Since we are interested in highly dynamic systems  the 
effects of velocity on the radiation field have to be considered 
(\cite{Ca72}, \cite{Mi78}). 
If all the relativistic terms to $O(v/c)$ are included
 and the presence of scattering is considered, the comoving frame transfer
 equation takes the form (\cite{MM84}):
\begin{eqnarray}
\label{te}
\nonumber
 \frac{1}{c} \frac{\partial I_\nu(r,\mu,t)}{\partial t}+\beta \frac{\partial 
I_\nu(r,\mu,t)}{\partial r}\\
\nonumber
+\frac{\mu}{r^2} \frac{\partial \left (r^2 I_\nu(r,\mu,t) \right)}{\partial r}+
    \\
\nonumber  
 \frac{\partial}{\partial \mu}
\left[ \left(1-\mu^2 \right) \left(\frac{1}{r}+\mu \left (\frac{\beta}{r}-
\frac {\partial \beta}{\partial r} \right) \right)
I_\nu(r,\mu,t) \right]
    \\ 
\nonumber
-\frac{\partial}{\partial \nu} \left [\nu \left(\left(1-\mu^2 \right)
\frac{\beta}{r}+{\mu^2}\frac{\partial \beta}{\partial r}\right)I_\nu(r,\mu,t)
\right]
   \\ 
\nonumber
 +\left( \left(3-\mu^2\right)\frac{\beta}{r}+
\left(1+\mu^2 \right) 
\frac{\partial \beta}{\partial r}\right) I_\nu(r,\mu,t)
    \\  
 =\chi_{\nu}\left(\epsilon_{\nu}B_\nu(T)-
\left(1-\epsilon_{\nu}\right)J_{\nu} \right)-
\chi_{\nu}I_\nu(r,\mu,t) 
\end{eqnarray}
with, 
\begin{equation}
\chi_{\nu}=\sigma_{\nu}+\kappa_{\nu}
\nonumber
\end{equation}
and,  
\begin{equation}
\epsilon_\nu=\frac{\kappa_\nu}{\chi_\nu}
\nonumber
\end{equation}
$\sigma_\nu$ and $\kappa_\nu$ are the scattering and pure absorption components
of the opacity and $\epsilon_{\nu}$ is defined as the pure absorption fraction.
The other quantities have their usual meaning. The 
scattering is assumed to be coherent and isotropic and its presence makes
the equation (\ref{te}) to become integro-differential due to the term
$J_{\nu}$.

Integrating (\ref{te}) over frequency and calculating its first and second
moments respect to $\mu$ it is possible to obtain the first and second
frequency integrated moment equations.  
\begin{equation}
\label{eq}
\begin{array}{l}
\displaystyle\frac{1}{c}\frac{\partial J}{\partial t}+\beta \frac{\partial J}{\partial r}+
\frac{1}{r^2} \frac{\partial \left ( r^2 H \right )}{\partial r} +
 \frac{\beta}{r} \left(3J-K \right) +\frac{\partial \beta}{\partial r}(J+K) \\
 \\
 \displaystyle= \left ( \chi_{\rm P}- \sigma_{\rm P} \right ) B(T)-
\chi_{\rm J}J+\sigma_{\rm J}J
\end{array}
\end{equation}
\vspace{1.5 cm}
\begin{equation}
\label{mq}
\begin{array}{l}
\displaystyle \frac{1}{c} \frac{\partial H}{\partial t}+ \beta 
\frac{\partial H}{\partial r}+
 \frac{\partial  K}{\partial r}+\frac{3K-J}{r}+2  
\left(\frac{\partial \beta}{\partial r} +\frac {\beta}{r} \right)H \\
\\
 \displaystyle  =-\chi_{\rm H}H 
\end{array}
\end{equation}

In (\ref{eq}) and (\ref{mq}) the ``mathematical'' radiation moments are
related to the ``physical'' ones just by
 $E=\frac{4  \pi}{\rm{c}}J$, $F=4 \pi H$ and  $P=\frac{4 \pi}{\rm{c}} K$. 
 Since
the equations are frequency integrated, the opacities 
($\chi_{H}$, $\chi_{J}$ and
 $\chi_{P}$) are mean opacities: the flux mean, the absorption mean and the Planck mean
respectively. In this work only grey opacities are considered
 and thus $\chi_{H}$=$\chi_{J}$=$\chi_{P}$.

Since the general moment equations (\ref{eq}) and (\ref{mq})contain three
 unknowns it is necessary to 
add a third equation:
\begin{equation} 
\label{gcr}
\Phi (J, H, K,r,t)=0
\end{equation}
This equation is known as the closure relationship and it recovers part
of the geometrical information lost after integrating (\ref{eq}) over $\mu$. 
Notice that the set formed by the equations (\ref{eq}), (\ref{mq}) and 
(\ref{gcr}) 
is equivalent to the diffusion approximation in the 
asymptotic limit for optically thick media:
\begin{equation}
\label{deq}
\begin{array}{l}
\displaystyle\frac{1}{c}\frac{\partial J}{\partial t}+\beta \frac{\partial J}{\partial r}+
\frac{1}{r^2} \frac{\partial \left ( r^2 H \right )}{\partial r} +
 \frac{\partial \beta}{\partial r}\frac{4}{3}J    
= \kappa_{\rm P}\left(B(T)-J \right )
\end{array}
\end{equation}
\begin{equation}
\label{dmq}
\chi_{\rm R} H=-\frac{1}{3}\frac{\partial J}{\partial r}
\end{equation}
Although the closure relationship has apparently disappeared, it is 
implicitly included. Furthermore, several velocity dependent terms as well as the 
scattering contribution have been dropped and the flux
is instantaneously determined by the energy density distribution since 
no time derivative appears in equation
 (\ref{dmq}). All these properties simplify the numerical treatment of the
problem, but its accuracy decreases as the optical depth of the medium 
decreases.
 
A  remark valid for both sets of equations is that all the radiation
moments appearing in them are comoving moments. For this reason they always 
have to be transformed to inertial frame moments to compare with observations.
 This is done by using the transformation equations (\cite{MM84}):
\begin{equation}
J'=J+2\beta(H)
\end{equation}
\begin{equation}
H'=H+\beta(J+K)
\end{equation}
\begin{equation}
K'=K+2\beta(H)
\end{equation}

There are two  cases where (\ref{gcr}) is
known a priori: planar media with grey
 opacity  and optically thick regions, where the radiation field is 
quasi-isotropic.
 In the former case an essentially exact, in the latter the particularly 
simple Eddington closure relationship (\ref{ecr}) holds.
\begin{equation}
\label{ecr}
3K-J=0
\end{equation}
When the anisotropies of the radiation field due to geometrical effects become
important, the equation (\ref{ecr}) is no longer valid. An example  
is the radiation field far away from a quasi-punctual source
which verifies $K-J \approx 0$ (Stone et al. 1992).
Nevertheless, due to its
simplicity,  equation (\ref{ecr}) is very often applied and, in fact, it is assumed
when using the diffusion approximation. 

Generally, when (\ref{ecr}) fails, the
``variable Eddington factor'' ($K/J=P/E= f (r,t)$) is used. The 
factor is computed solving from time to time the stationary transfer
equation and it is assumed to be constant between successive determinations. 
This method implies an important increase in the complexity and time 
consumption of the calculations.

Simonneau (1979) proposed a generalization of equation (\ref{ecr}) for 
spherically symmetric systems where radiation anisotropies are important. 
In the grey case, the equation takes the form:
\begin{equation}
\label{simcr}
3K-J=2 \mu_{c}~ H \:\:\:\:\:\:\:\: \left ( 0 \le 
\mu_{c} \le 1 \right )
\end{equation}
where the parameter $\mu(r,t)_{c}$ is given by: 
\begin{equation}
\label{simeq}
\frac{d\mu_{c}}{d r}=\frac{1-\mu_{c}}{r \mu_{c}}
 \rm{e}^{-\frac{\tau}{\mu_{c}}}\:\:\:\:\:\:\: \mu_{c} \left( 0 \right) = 0
\end{equation}
It reproduces the equation (\ref{ecr}) ($\mu_{c}$=0) and the streaming limit
($\mu_{c}$=1) as extreme cases. 
In  previous papers (Lopez et al. 1987, Simonneau et al. 1989) it was proven 
the efficiency of (\ref{simeq}) and (\ref{simcr}) to produce a suitable closure relationship
 for static and stationary envelopes.

However, high velocities  produce aberration and Doppler shift effects
 on the travelling photons and they might modify the form of  this
closure relationship respect to the the static case.
 With the purpose of evaluating such  effects, several stationary but 
non-static envelopes with  suitable velocity fields have been examined.
The radiation field of these systems was  determined by  
solving the complete transfer equation with and without a velocity field.
  In the former case the moments were obtained, both in the
  comoving-frame and the inertial-frame. Values of $\mu_{c}$ were
 computed using  (\ref{simcr}).
 Figure (\ref{nescr}) displays   the profiles obtained for one of these
envelopes. 
 In all  the cases the three profiles were similar. In particular, the 
 differences  between
the static value and the non-static one in the inertial frame are negligible. 
The conclusion is that the 
 closure relationship for the inertial-frame moments of the radiation field 
 was not noticeably
affected by the presence of velocity fields (at least, in the
moderately-relativistic case). The reason for that is that the values adopted
 by  $\mu_{c}$ are only significant (i.e. $\mu_{c} \not \ll 1$) in low 
opacity regions, where the radiation is not strongly coupled to the local
conditions and 
velocities do not have a
significant influence on the photon distribution.

\placefigure{nescr}

Once it is  assumed the invariance of the closure relationship for the 
inertial-frame moments, it is possible to write in the comoving-frame:
\begin{equation}
\label{jorcr}
3K \left ( 1+\frac{2}{3}\mu_{c}\beta \right )- J \left (1 + 
2 \mu_{c} \right) = 2 \mu_{c} H \left ( 1 + 2 \frac{\beta}
{\mu_{c}} \right )
\end{equation}
Which is the final expression we used in this work.
	
On the other hand, it is also necessary to estimate the validity of the 
hypothesis of stationarity on the expression (\ref{jorcr}).
The necessary and sufficient condition for its applicability in dynamic 
situations is:
\begin{equation}
\tau_{\rm{radiation}} \ll \tau_{\rm{evolution}}
\end{equation}
where $\tau_{\rm{radiation}}$ is the characteristic time for the evolution
 of the radiation field and $\tau_{\rm{evolution}}$ is the characteristic
time for the evolution of the system. In  regions where the role of
 (\ref{jorcr}) is relevant (optically thin), $\tau_{\rm{radiation}}$  
corresponds to the free flight time (\cite{MM84}), which 
 is generally  shorter than the evolution time for any fluid flow system.
 In particular,  in the case of  SNIa, $\tau_{\rm{evolution}}$
 is determined by the dynamical time and the  decay time of
 $^{56}$Ni and $^{56}$Co, which are indeed much longer than 
 $\tau_{\rm{radiation}}$.
The use of (\ref{jorcr}) and (\ref{simeq}) is then justified. 

The calculation of a model of supernova envelope involves not only the 
integration of the equations describing the radiation field but also 
those describing the evolution of matter.
Fortunately, there are several properties of
supernovae that make possible to simplify the problem: i) During the 
homologous
expansion, the structure of the envelope can be easily determined  
if the initial velocity profile is known. ii) Due to the low densities
and high temperatures, the energy content of the envelope is dominated
by radiation, and since radiative balance of the gas can be assumed, 
it is possible
to replace the internal energy equation by an equilibrium condition. With
these assumptions and using the general moment equations, the expressions 
which describe  the evolution of the
energy density take the form (now in Lagrangian coordinates):
\begin{equation}
\label{aja}
\frac {1}{c} \frac {\partial J}{\partial t}  + 4 \pi \rho \frac
{\partial \left ( r^2 H \right)}
{\partial M}
-  \beta \frac {\left ( 3K -J \right )}{r} -\frac{\left ( J+K \right ) }
{c} \frac {\partial \rho}{\partial t} \frac {1}{\rho}=\frac {\xi \rho}{4 \pi}
\end{equation}
\begin{equation}
\label{ajab}
 4 \pi \epsilon \chi \left (B(T_{\rm{gas}})-J \right )=\xi \rho
\end{equation}
 Alternatively, if the diffusion approximation
is considered (and hence Eq. \ref{deq}), the corresponding equations
will be:
\begin{equation}
\frac {1}{c} \frac {\partial J}{\partial t}+ 4 \pi \rho \frac
{\partial \left (r^2 H \right )}
{\partial M} -\frac {4}{3} \frac {J}{c} \frac {\partial \rho}{\partial t
}
\frac {1}{\rho}=
\frac {\xi \rho} {4 \pi}
\end{equation}
\begin{equation}
 T_{gas}=T_{rad}
\end{equation}
where $\xi$ is the radioactive energy deposition function.  For simplicity the
gas was considered to be in LTE, although a really  accurate description of
 the problem, which is beyond the scope of this work, would require a NLTE
 modeling (\cite{Ba95}). Hence, the ionization degree was computed
with the Saha equations. In fact, in the calculations this quantity
was from the tables previously previously obtained by Bravo et al. (1993).
The resulting set of differential equations has been solved using an implicit
scheme. The finite difference equations were space and time centered and 
second order accurate. The complexity of both sets of equations
 (diffusion approximation and moments) is similar and we have found that 
 their time consumption is almost similar. 

In order to verify the quality of the different approximations, we compare
the results obtained with them with those provided by the variable
Eddington factor method which, as in other works, is assumed to give
 correct results (\cite{St92}, \cite{Bl93},  \cite{Ho93}, \cite{En94}).
 In this case in order to compute the variable Eddington factor,
 the transfer equation is solved 
periodically in its stationary form with a second order accuracy scheme,
following the parallel ray technique (\cite{MM84}).

Besides the evolution of the bolometric quantities: J, H and K
, we have also computed the color temperature, T$_{UB}$, and the U, B 
and V light curves. To do that,  the stationary monochromatic
moment equations have been solved at fixed frequencies and times using the
values of  J and T$_{gas}$ obtained with the dynamic frequency-integrated
equations and introducing a correction 
factor to make them consistent with the previously computed values of H.

\section{Results and discussion}

The bolometric light curves obtained with the different approximations are 
displayed in Figure \ref{lc1} in one of the cases considered here while
Table \ref{table} summarizes the main differences among them. In all the 
cases, the generalized Eddington factor gives excellent results. At the 
maximum, the light curve is slightly underluminous and the differences are in 
the range of 0.02$^{m}$--0.05$^{m}$. In contrast the light curve obtained
with the diffusion approximation is clearly overluminous near the maximum 
although these discrepancies disappear a week or two later, when the 
luminosity balance is reached. These discrepancies range
from -0.14$^m$ to -0.2$^{m}$, being more important in  the case of models 
with lower optical depths at maximum, although this
dependence is not very strong. The errors in this case are not affected by 
the adopted value of $\epsilon$. The diffusion approximation also modifies
the position of the maximum luminosity  which occurs 1--2 days too early. See
Table \ref{table}

\placefigure{lc1}

Deviations of E$_{rad} \left ( r \right )$ =$4 \pi /c$ J$ \left ( r
 \right )$
 and of T$_{gas} \left ( r \right )$ are also very small for the 
proposed method. During the first 3 months the errors for these quantities
in the innermost and outermost layers are negligible. They  appear only in the
intermediate region, where they fluctuate above and below the correct
values (Fig. \ref{perfi3}). In this case the deviations do not disappear
at late times but grow as the models evolve.  During all this period,
relative discrepancies are less than 15\% and 4\% for E$_{rad}$(r) and 
  T$_{gas}$(r) respectively. 

\placetable{table}

For the same  models the discrepancies introduced by the
diffusion approximation appear at the surface just after the maximum luminosity 
 and  follow the recession of the photosphere towards inner regions. 
The estimated values
for E$_{rad}$ and T$_{gas}$ are always below the expected ones and their
profiles are too smooth. Two months after the explosion E$_{rad}$ is on average
 $\approx$ 50 \% of the correct value 
and T$_{gas}$ $\approx$30\% and the
simulated envelopes become quasi-isothermal, in opposition to those obtained
with the variable Eddington factor method. When the optical depth of
the models is close to 1 the diffusion approximation is unable to produce
useful values for these quantities.

\placefigure{perfi3}
\placefigure{col13}

The deviations of E$_{rad}$ and T$_{gas}$ affect the computation of the 
monochromatic light curves, the color temperature and the B--V index.
 The excessively low values of E$_{rad}$ given by the diffusion approximation
cause a redshift of the estimated spectra when models become transparent,
This produces a decrease of the color temperature and
an increase of B--V. T$_{UB}$ displays moderate deviations in the region 
of the maximum luminosity
 (-700 K -- 100 K) but, when the total optical depth of the models is $ <$ 2
it steeply falls to very low values as it is displayed in Figure \ref{col13}.
Similarly B--V evolves close to the correct behaviour during the first
weeks but it takes too high values as soon as T$_{UB}$ starts decreasing
 (Fig. \ref{bv}).
 The  U, B and V light curves  are influenced by the combination of the
spectral redshift and the overluminosity  at maximum caused by
the diffusion approximation. Near the maximum of luminosity, the second effect 
compensates or even dominates the spectral redshift, but  a month
later, even  the model with the highest opacity is underluminous in
 the U, B and V bands (Figure \ref{mcl1}). At 40 days, the
deviations  range from 0.2$^m$ to 3$^m$. All these discrepancies 
are sensitive not only to the total opacity, but also to the
pure absorption fraction. See Table \ref{table}

\placefigure{mcl1}
\placefigure{bv}

Once again, the accuracy provided by the generalized Eddington factor
for these quantities
is excellent. Even 3 to 4 months after the explosion T$_{UB}$, B--V and
the U, B and V light curves evolve very close to the ``standard'' values 
obtained
with the variable Eddington factor method. In all the models
the deviations from the standard results are negligible near the maximum of
luminosity. Only
after two months it is appreciated a moderate shift towards the blue. 
During the first 4 months $\Delta$T$_{UB}\le$ 100
 and the B-V color excess is $\approx$ -0.05$^{m}$. For the same interval
the absolute errors of  the monochromatic light curves are always below
 0.1$^m$. These errors do not show a clear dependence on the 
opacity as it can be seen in Table \ref{table}. This is a consequence of
 the ability of the method to handle systems even with very low 
optical depth.  The errors in this case depend on the pure absorption fraction
 $\epsilon$ of the model. 

\section{Conclusions}
The possibility of applying to dynamical situations the closure relationship
proposed by Simonneau (1979) for extended envelopes has been verified.
 The use of a slightly modified version of this closure relationship
 together with the comoving frame moment equations is a suitable
method for solving the radiative transfer problem. This method takes into
account velocity terms as well as scattering effects, and allows the
simultaneous treatment of the optically thin and thick regions, giving 
much more accurate
results than those obtained with the diffusion approximation. 
Since the closure
relationship can be determined a priori, no significative complexity is added
to the calculations and the computational effort is similar to that 
required by the diffusion approximation.

Our results show that the method provides very precise results for all
the quantities in all the scenarios considered here, even at the late 
epochs when the envelope becomes transparent. For
the same models the diffusion approximation fails in reproducing all the
properties other than bolometric luminosity a month after the explosion.

{\em Acknowledgements}. We are grateful to E. Simonneau for helpful comments
 concerning this paper and to R. L\'opez for information on her previous work
in this subject. This work has been financed with the CICYT project
 ESP95-0091.

%
%

\newpage
\begin{deluxetable}{ccccccccc}
\tablecaption{DEVIATIONS FROM THE STANDARD MODEL OF DIFFERENT APPROXIMATIONS 
\label{table}\tablenotemark{a}}
\tablenotetext{a}
 {(GEC) corresponds to simulations performed with the Simonneau's 
generalization of the Eddington closure
  and (DAM) to the simulations performed with the diffusion approximation.}
\tablehead{
\colhead{} &
\colhead{GEC} &
\colhead{DAM} &
\colhead{GEC} &
\colhead{DAM} &
\colhead{GEC} &
\colhead{DAM} &
\colhead{GEC} &
\colhead{DAM} \nl
\colhead{\large{$\chi, \epsilon$}} &
\multicolumn{2}{c}{($\Delta$T$_{UB}$)$_{max}$} &
\multicolumn{2}{c}{($\Delta$M$_{bol}$)$_{max}$} &
\multicolumn{2}{c}{($\Delta$V)$_{40}$} &
\multicolumn{2}{c}{($\Delta$B)$_{40}$}}
\startdata
0.2$\rho$, 1 & -110 & -300  & 0.03 & -0.14 & 0.02 & 0.2 & 0.05 & 0.18
\nl
~~~~~~~~, 0.01&  -70 & 100 & 0.03 & -0.14 & -0.01 & 0.4 & -0.02 & 0.9 
\nl
0.05$\rho$, 1 & -50 & -700  & 0.02 & -0.16 & 0.01 & 0.6 & 0.01 & 0.8 
\nl
~~~~~~~~, 0.1& -50 & -200 & 0.02 & -0.16 & 0.02 & 0.6 & 0.02 & 0.8 
\nl
~~~~~~~~, 0.01&  -80 & -100 & 0.01 & -0.16  & -0.02 & 1.6 & -0.03 & 2.1 
\nl
$\chi(\rho,T)$ , 1 &  -50 & -600  & 0.01 &-0.2 & -0.06 &2.2 & -0.02 & 3.1 
\nl
~~~~~~~~, 0.1& -30 & 80 & 0.02& 0.2 & -0.03 & 3.2 & -0.01 & 3.8 
\nl
\enddata
\nl
\end{deluxetable}

%
%

%
%

\newpage
\figcaption[fig1.eps]{
Values of $\mu_c$ for a SNIa 30 days after the explosion. Solid line
corresponds to the static case, dashed line and dotted line correspond to
the values obtained in the non static case with the comoving-frame and 
inertial-frame values respectively.\label{nescr}}
\figcaption[fig2.ps]{
Bolometric light curves for the model with temperature dependent 
opacity  and pure absorption fraction $\epsilon=1$. The light curves have
been obtained with the diffusion approximation (dotted line), the proposed
method (dashed line) and the variable Eddington factor method (solid line).\label{lc1}}

	\figcaption[fig3.eps]{
E$_{rad}$ evolution for models with  $\chi=$ 0.2 $\rho$ (top) and
 $\epsilon$=1,  $\chi$=0.05 $\rho$ and $\epsilon=1$ (bottom). Lines have the
usual meaning.\label{perfi3}}
	\figcaption[fig4.eps]
{T$_{color}$ as a function fo the time for the models with: (a)
 $\chi$=0.05 $\rho$  and $\epsilon=1$, (b) $\chi$=$\chi(\rho,T)$ and 
$\epsilon$=1. Curve (b) is 3000 K below its actual position. Lines have
the usual meaning.\label{col13}}
	\figcaption[fig5.eps]
{V and B (shifted 1 magnitude) light curves. Model with $\chi$=0.05 $\rho$ 
and $\epsilon=1$. Lines have the usual meaning.\label{mcl1}}

\figcaption[fig6.eps]
{Evolution of the color index B--V for  models with $\chi$=
0.05 $\rho$. (a) $\epsilon$=1, (b) $\epsilon$=0.1. Curves of model (a)
have been shifted 1 magnitude. Lines have their usual meaning.
\label{bv} }

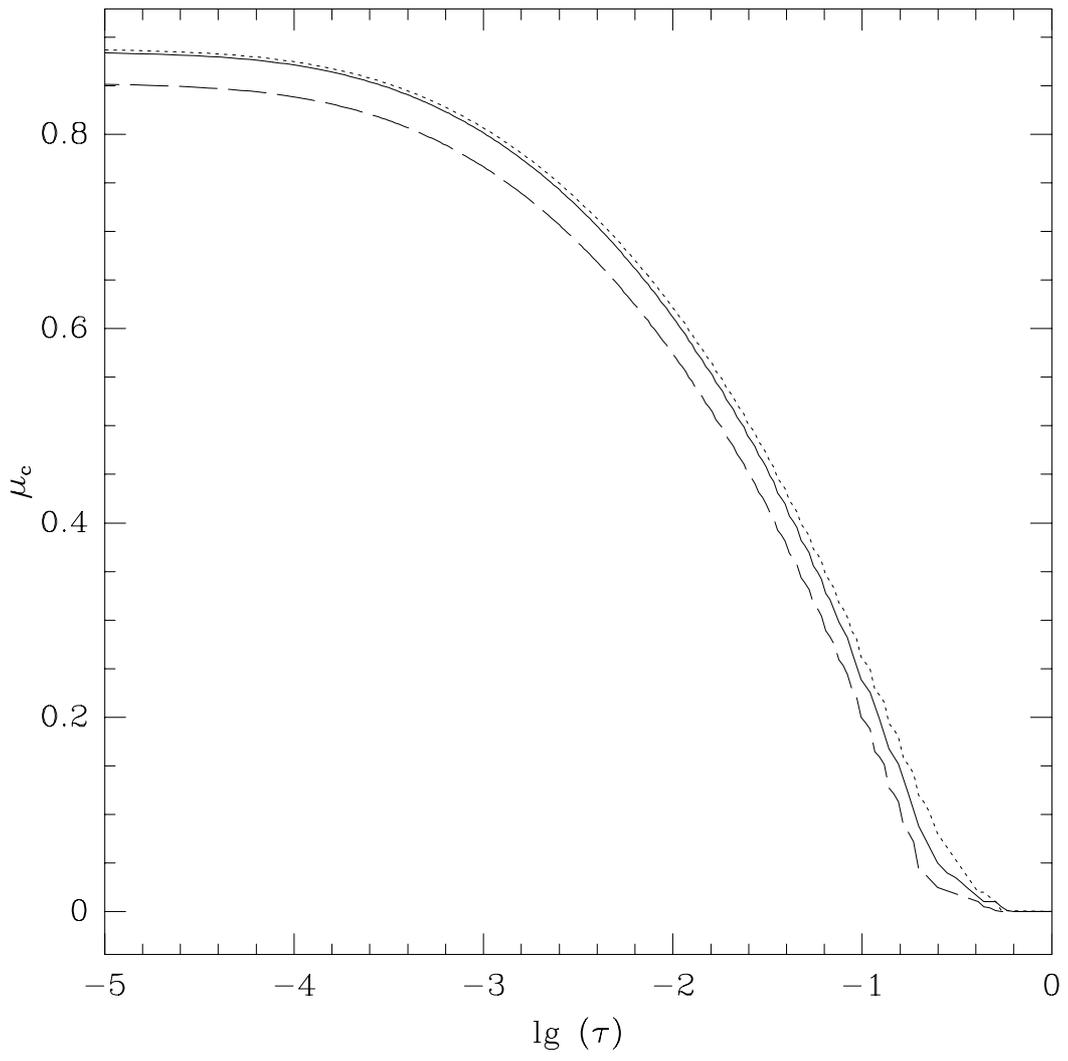
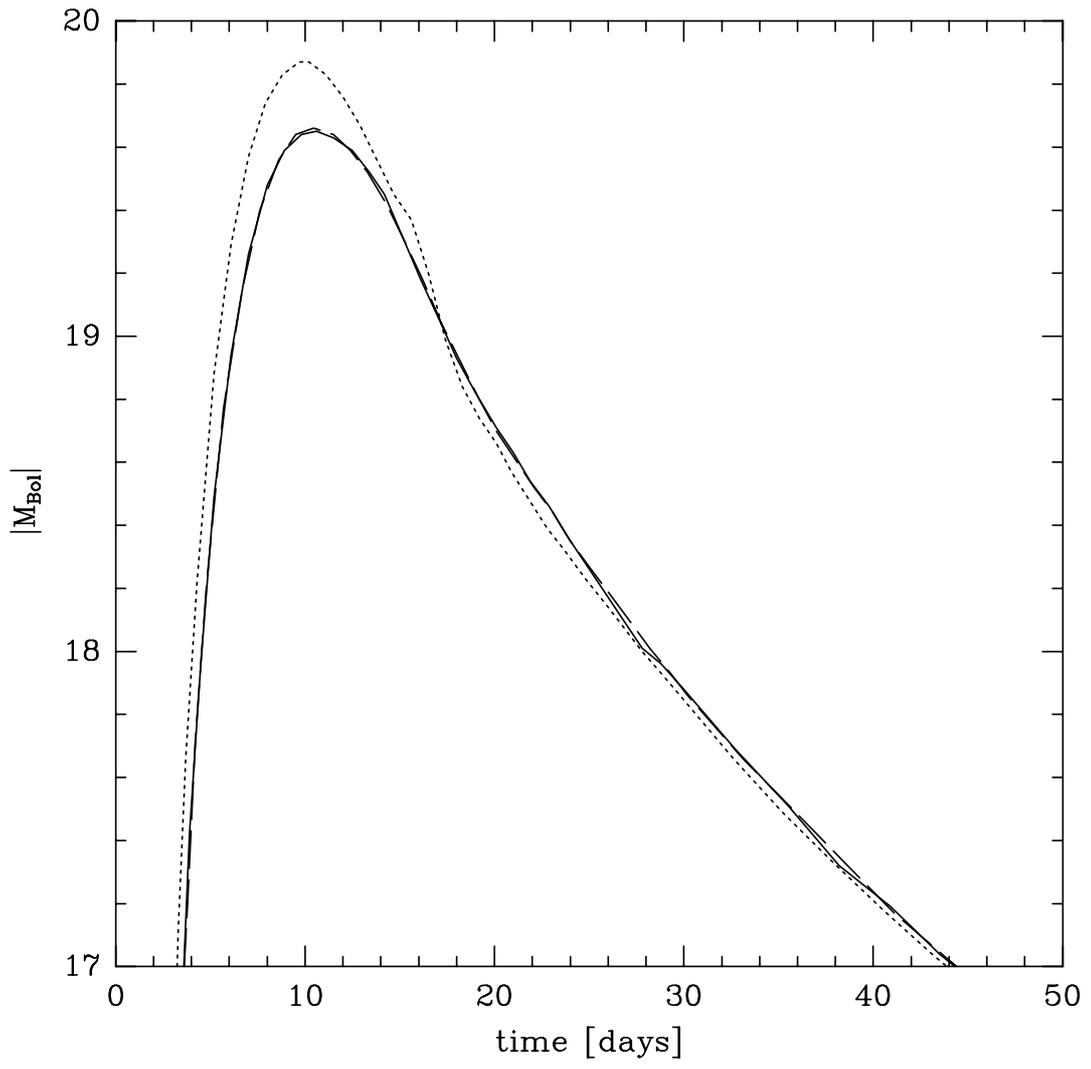
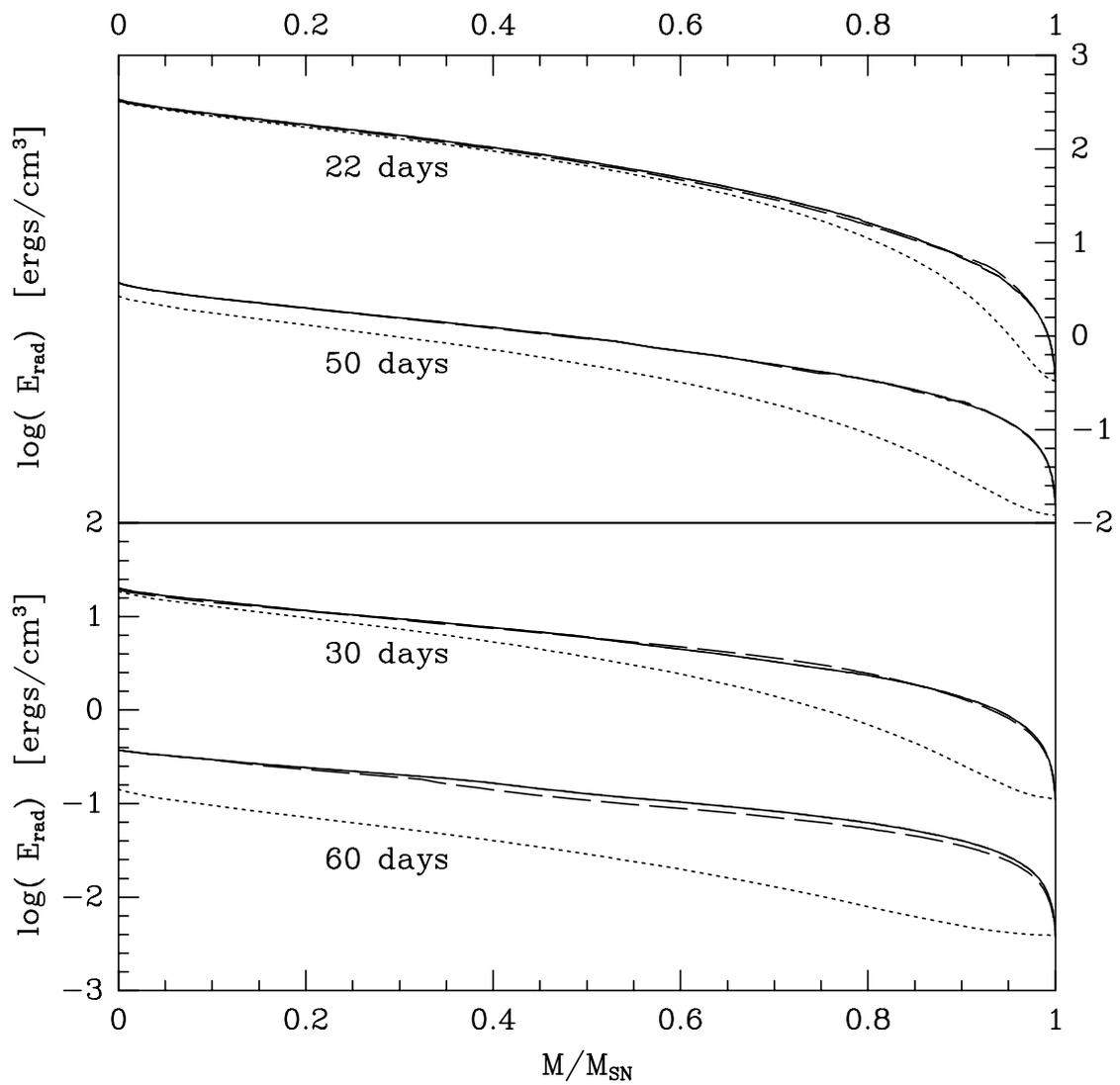
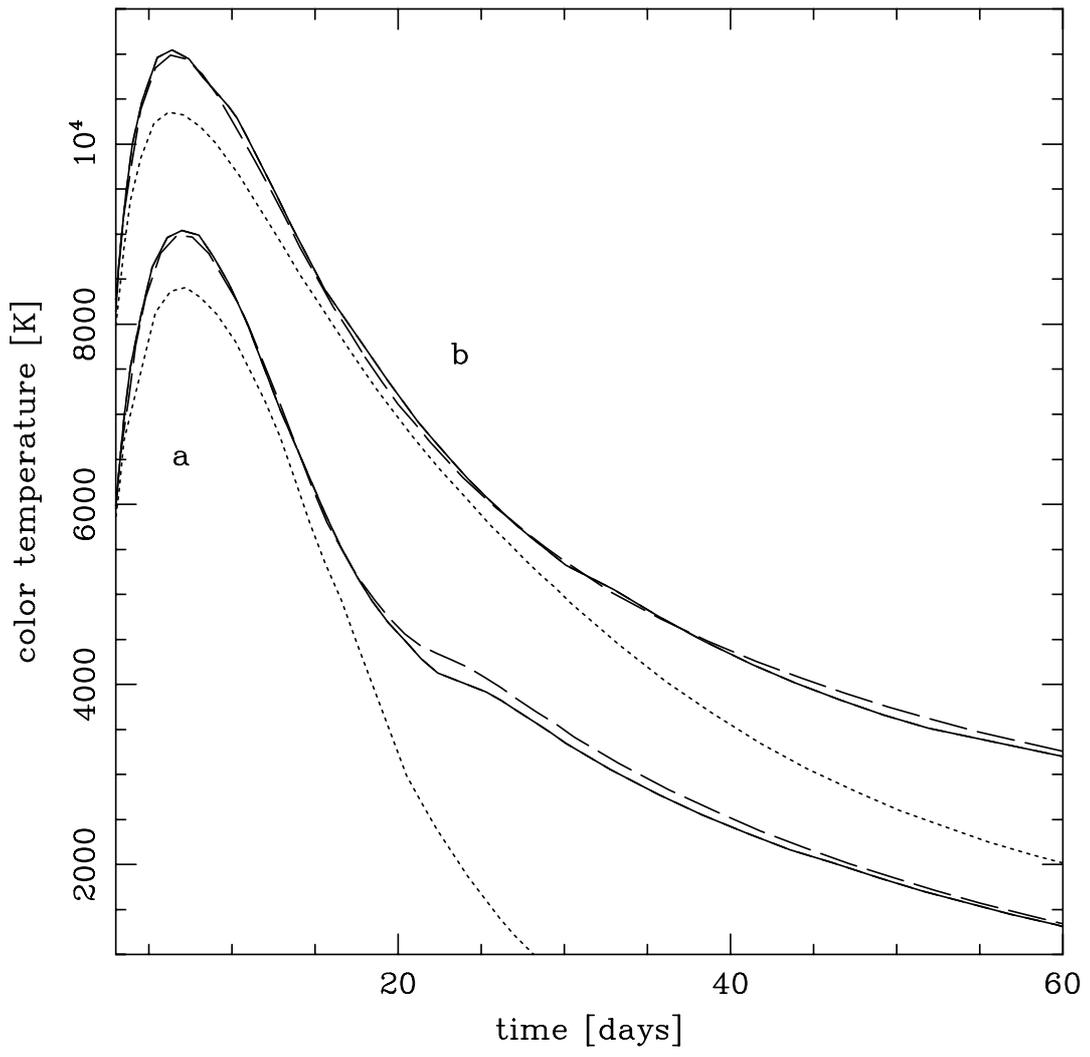
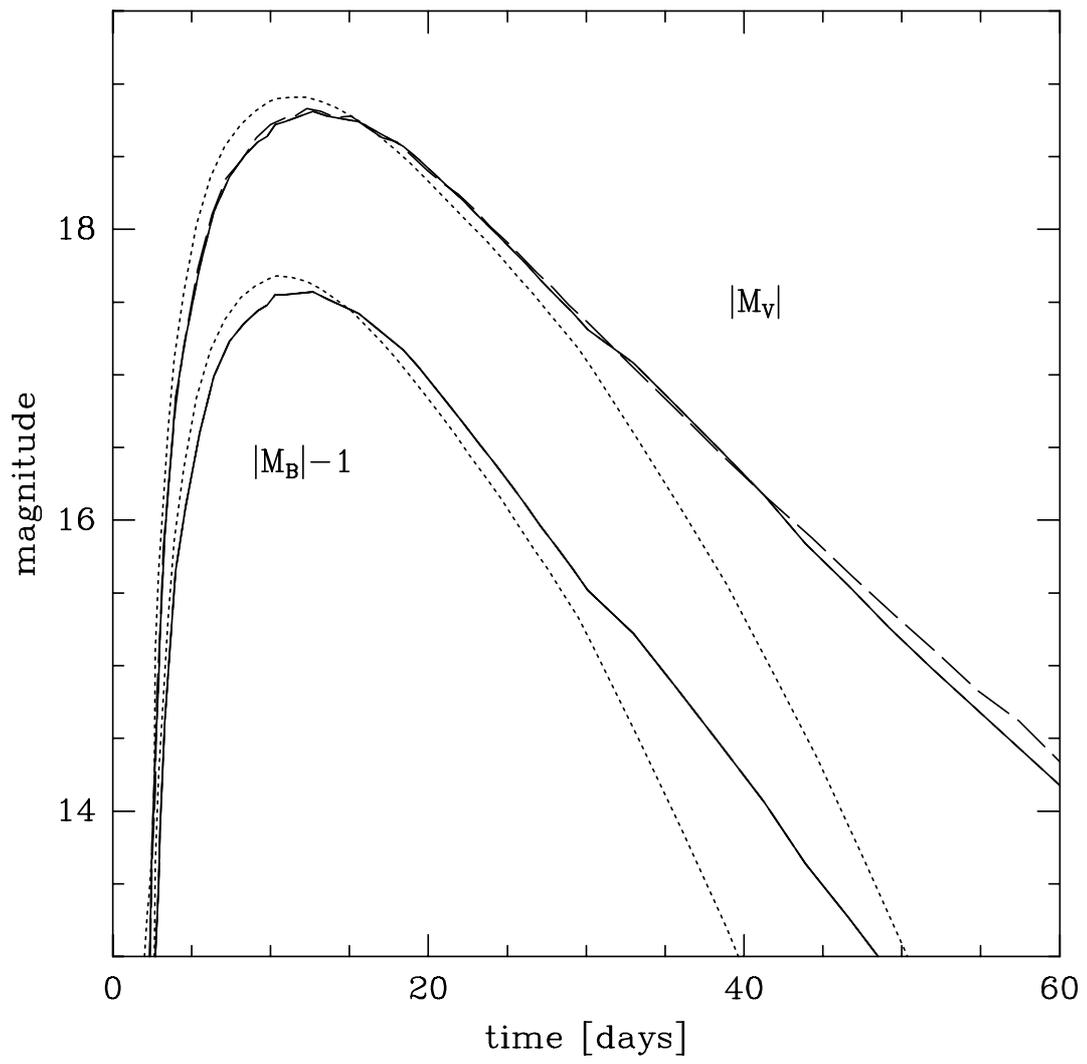
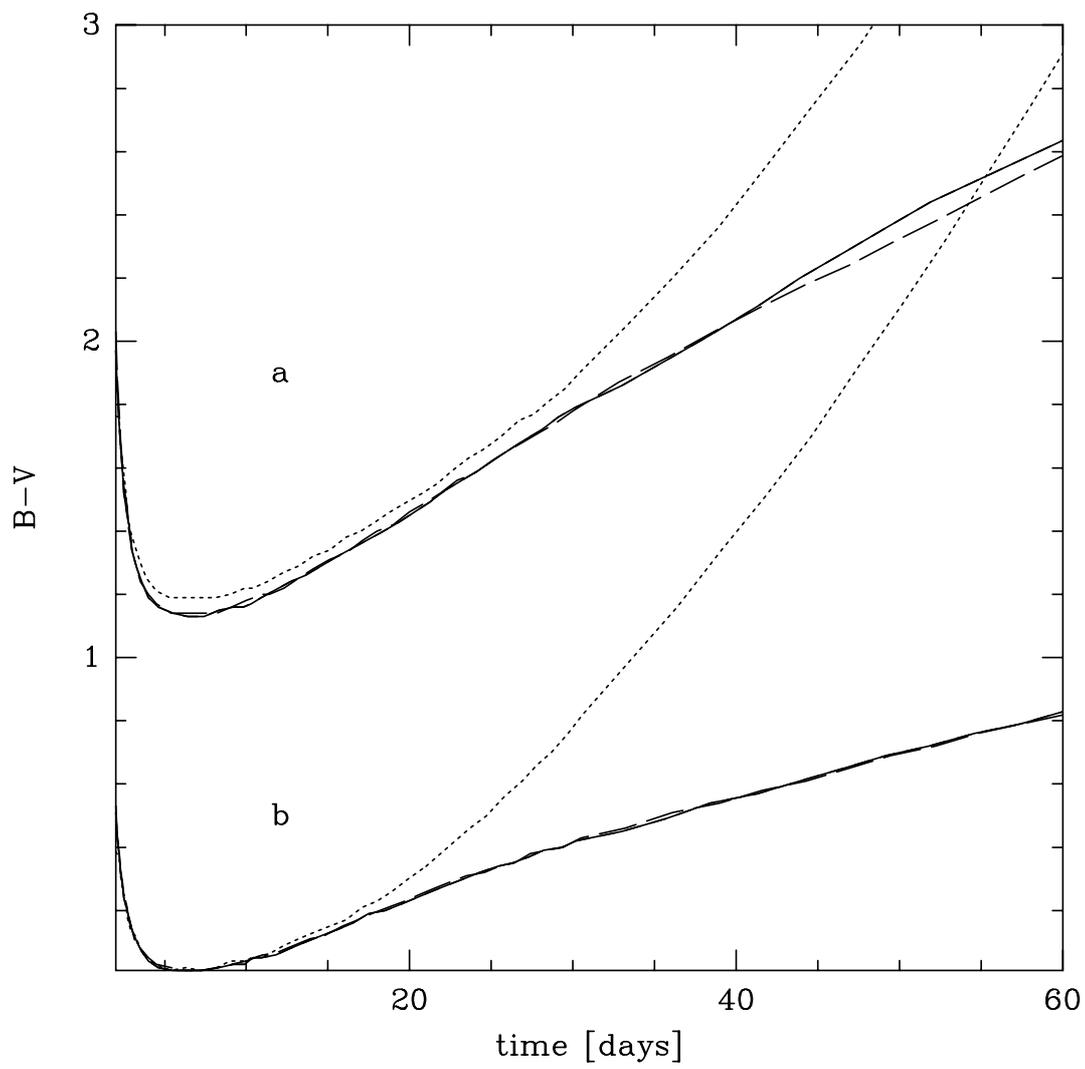
\end{document}